\documentclass[twocolumn,showpacs,preprintnumbers,amsmath,amssymb,floatfix]{revtex4}
\usepackage{graphicx}
\usepackage{dcolumn}
\usepackage{bm}
\begin{document}
\title{A crystallographic phase transition within the magnetically ordered state of Ce$_{2}$Fe$_{17}$}
\author{A. Kreyssig}
 \altaffiliation[Also at ]{Institut f\"ur Festk\"orperphysik, Technische Universit\"at Dresden, D-01062 Dresden, Germany}
 \email{kreyssig@ameslab.gov}
\author{S. Chang}
\author{Y. Janssen}
\author{J.-W. Kim}
\author{S. Nandi}
\author{J. Q. Yan}
\author{L. Tan}
\author{R. J. McQueeney}
\author{P. C. Canfield}
\author{A. I. Goldman}
 \affiliation{Ames Laboratory USDOE and the Department of Physics and Astronomy, Iowa State University,
Ames, IA 50011, USA}%
\date{\today}

\begin{abstract}
X-ray diffraction experiments were performed on polycrystalline 
and single-crystal specimens of Ce$_{2}$Fe$_{17}$ at temperatures 
between 10~K and 300~K. Below $T_{\mathrm{t}}$~=~118$\pm$2~K, 
additional weak superstructure reflections were observed in the 
antiferromagnetically ordered state. The superstructure can be 
described by a doubling of the chemical unit cell along the 
$\mathbf{c}$ direction in hexagonal notation with the same space 
group $R~\overline{3}~m$ as the room-temperature structure. The 
additional antiferromagnetic satellite reflections observed in 
earlier neutron diffraction experiments can be conclusively 
related to the appearance of this superstructure.
\end{abstract}

\maketitle{}

\section{Introduction}

The magnetic states in $R_{2}$Fe$_{17}$ compounds ($R$~= rare 
earth) are mainly determined by the exchange interactions between 
the Fe moments,\cite{Buschow77} where competition between 
antiferromagnetic and ferromagnetic interactions results in the 
appearance of an antiferromagnetic or ferromagnetic ground state 
related to the $R$ ion's moment, anisotropy and/or 
size.\cite{Coey93,Makihara03} The balance between these 
interactions seems particularly delicate for Ce$_{2}$Fe$_{17}$, 
where even slight structural or chemical changes can strongly 
modify the magnetic behavior.\cite{Janssen97}

Ce$_{2}$Fe$_{17}$ orders antiferromagnetically below the N\'eel 
temperature $T_{\mathrm{N}}\sim208$~K. At 
$T_{\mathrm{t}}\sim125$~K a second transition is observed into a 
modified antiferromagnetic state.\cite{Janssen97} Several 
publications have also reported a ferromagnetic state for their 
samples at low 
temperatures,\cite{Makihara03,Kuchin00,Prokhnenko02a} but this is 
most likely caused by small amounts of dopants in the 
investigated samples.\cite{Janssen06} This interpretation is 
supported by systematic studies of series of samples with partial 
substitution of Fe by Al, Si or 
Mn.\cite{Mishra96,Artigas98,Kuchin00,Kuchin00a,Pirogov00} 
Furthermore, a Ce$_{2}$Fe$_{17}$ sample with a ferromagnetic 
state at $T$~=~40~K at ambient pressure shows a transition into 
an antiferromagnetic order under isostatic compression at 
3~kbar.\cite{Prokhnenko02a,Prokhnenko02,Prokhnenko04} These 
studies indicate that the magnetic behavior is extremely 
sensitive to slight changes of the crystal structure either 
through modifications of the chemical composition or 
thermodynamic parameters, like pressure. Significant 
magneto-elastic coupling has also been observed in the course of 
a study of the magnetic phase diagram where it was noticed that 
some phase transitions, in an applied magnetic field, were 
associated with dramatic changes in the shape of the 
sample.\cite{Janssen06} So far, the relationship between the 
magnetic behavior and changes in the crystal structure in 
Ce$_{2}$Fe$_{17}$ have been mainly based on analysis of lattice 
parameters or the volume of the crystallographic unit cell. Here, 
we present an x-ray investigation on powder and single crystals 
focussed on qualitative changes in the crystal structure of 
Ce$_{2}$Fe$_{17}$.

Ce$_{2}$Fe$_{17}$ crystallizes in the Th$_{2}$Zn$_{17}$-type 
structure with the rhombohedral space group $R~\overline{3}~m$ 
(No.~166).\cite{Buschow77} In the following, the hexagonal 
description is used. At room temperature the lattice parameters 
are $a$~=~8.4890~\AA~ and $c$~=~12.410~\AA. The Ce atoms are 
located on the Wyckoff site 6$c$ (0,~0,~0.3435). The Fe atoms 
occupy four different sites, 6$c$ (0,~0,~0.0968), 9$d$ 
(1/2,~0,~1/2), 18$f$ (0.2905,~0,~0), and 18$h$ 
(0.5015,~0.4985,~0.1550).\cite{Janssen06} In two independent 
neutron diffraction investigations, satellite reflections were 
observed below $T_{\mathrm{N}}$ related to the antiferromagnetic 
structure with an incommensurate propagation vector in the 
hexagonal $\mathbf{c}$ direction. Below $T_{\mathrm{t}}$ 
additional magnetic satellite reflections were present. 
Interestingly, these reflections were indexed as 
(0~0~3/2)$^{\pm}$ identifying them as satellites to a reference 
vector (0~0~3/2).\cite{Plumier74,Fukuda99,Fukuda99a} The authors 
concluded that the incommensurate antiferromagnetic structure 
``is derived from a superstructure magnetic cell with a $c$ 
parameter twice as large as the one of the nuclear 
cell''.\cite{Plumier74} However, in all scattering studies to 
date no reflections at the position of the reference vector 
(0~0~3/2) have been reported.

In the present investigation, these corresponding superstructure
reflections have been observed in well-characterized
Ce$_{2}$Fe$_{17}$ samples. The superstructure reflections related 
to (0~0~3/2) were studied as function of temperature between 10~K 
and 300~K. Furthermore, entire reciprocal planes were measured, 
using high-energy x-rays and image-plate technology, in order to 
identify additional superstructure reflections. This 
superstructure will be discussed in terms of its relation to the 
room-temperature structure and its interactions with the magnetic 
and electronic properties.\\

\section{Sample Preparation and Experimental Conditions}

Ce$_{2}$Fe$_{17}$ samples were grown by a self-flux 
technique.\cite{Janssen06} Two types of samples were prepared for 
the present investigations. Single crystals with shiny surfaces 
were prepared in the same manner as the samples investigated in 
Ref.~\cite{Janssen06}. For the x-ray powder diffraction 
measurements, a polycrystalline sample was prepared by grinding 
one single crystal into fine powder which was kept under acetone 
to prevent oxidation. In order to achieve a uniform x-ray path 
through the sample, this polycrystalline material was pressed 
into the hole of a 3~mm diameter copper disk that measured 1.2~mm 
thick. Both sides of the disk were covered with kapton foil to 
keep the powder in place.  For single-crystal x-ray diffraction 
experiments, a plate-like single crystal with dimensions 
3~x~2~x~0.5~mm$^{3}$ was selected and attached to a flat copper 
sample holder. The same single crystal was used in all 
single-crystal measurements.

For both the x-ray powder and single crystal diffraction
measurements, the samples were mounted on the cold finger of a
helium closed-cycle cryostat where the temperature can be varied
between 10~K and 300~K. Two configurations for thermal shielding
were employed in these measurements.  First, three beryllium domes
were used for thermal isolation. The inner dome was filled with
helium gas to ensure good thermal exchange. A second dome between 
the inner and the outer domes provided additional heat shielding. 
The outer dome served as a vacuum enclosure. However, even though 
the elastic scattering of x-rays by beryllium is weak, the 
strongly textured Debye-Scherrer rings were found to impair the 
observation of very weak reflections from the sample. Therefore, 
during the search for superstructure reflections, an aluminum can 
with kapton windows was used. Although the sample temperature was 
no longer as well defined, the sample enclosure produced only a 
weak diffuse background. To define the sample temperature, 
measurements were performed in both configurations by keeping the 
sample in place and exchanging only the sample enclosures.

Two x-ray scattering instruments were employed. Preliminary single
crystal measurements were made on a standard four-circle
diffractometer using Cu-$K\alpha$ radiation from a high intensity
rotating anode x-ray source, selected by a silicon (1~1~1)
monochromator. The synchrotron x-ray diffraction measurements on 
the powder and single crystal samples were performed using a 
six-circle diffractometer at the 6-ID-D station in the MU-CAT 
sector at the Advanced Photon Source, Argonne. The synchrotron 
radiation, with a selected energy of 88~keV, allowed 
investigations of bulk samples because the absorption length is 
approximately 2.2~mm for Ce$_{2}$Fe$_{17}$, larger than the 
thickness of either the polycrystalline or single crystal 
samples. The beam size was 500~x~500~$\mu$m$^{2}$ limited by a 
rectangular slit system. For the high-energy x-ray measurements 
of both the powder and single crystal samples, the full 
two-dimensional diffraction patterns were recorded using a MAR345 
image-plate system positioned 730~mm behind the sample. The 
active detection area of 3450~x~3450 pixels of size 
100~x~100~$\mu$m$^{2}$ covered a total scattering angle 2$\theta$ 
of 13~deg, or a reciprocal space range of 12~{\AA}$^{-1}$ at this 
energy.

\section{Measurements and Results}

\subsection{Laboratory x-ray measurements of a single crystal}

In the preliminary diffraction study of a Ce$_{2}$Fe$_{17}$ 
single crystal on a standard four-circle diffractometer, strong 
reflections were found at positions ($H~K~L$) in hexagonal 
notation with $-H~+~K~+~L~=~3n$ where $H$, $K$, $L$, and $n$ 
integers according to the space group $R~\overline{3}~m$ for the 
known room-temperature crystal structure. In what follows, we 
categorize these reflections as the main reflections. Below 
$T_{\mathrm{t}}$, additional reflections were systematically 
observed at positions ($H~K~L+3/2$), as shown in FIG.~1. Even 
though these additional reflections are weak, they indicate an 
enlarged unit cell, or crystallographic superstructure. Hence 
these reflections are categorized superstructure reflections. 
Their measured intensity is  nearly four orders of magnitude 
lower than that of the main reflections, as illustrated in 
FIG.~1. The transverse peak widths of both sets of reflections 
(rocking curves), however, are similar (0.03~deg). While these 
measurements clearly establish the existence of a 
crystallographic superstructure in Ce$_{2}$Fe$_{17}$ below 
$T_{\mathrm{t}}$, elucidation of the nature of the modified unit 
cell requires more extensive measurements of the superstructure 
diffraction peaks.

\begin{figure}
\includegraphics[width=0.8\linewidth]{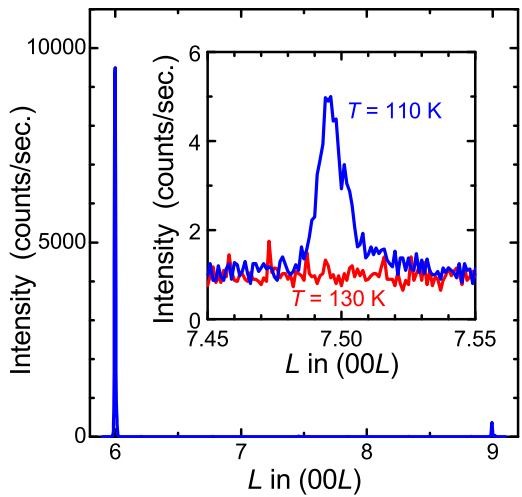}
\caption{\label{fig:fig1} (in color only online) Diffraction 
patterns of the Ce$_{2}$Fe$_{17}$ single crystal measured using 
Cu-$K\alpha$ x-rays at $T$~=~110~K and 130~K, well below and 
above the crystallographic phase transition, respectively. Parts 
of the patterns around the (0~0~15/2) position are enlarged in 
the inset.}
\end{figure}

\subsection{High-energy x-ray diffraction measurements of a powder
sample}

In an attempt to collect additional superstructure peaks 
associated with the modified chemical unit cell below 
$T_{\mathrm{t}}$, diffraction patterns of the polycrystalline 
Ce$_{2}$Fe$_{17}$ sample were measured at different temperatures 
between $T$~=~10~K and 300~K using the high-energy x-rays 
available in Station 6ID-D of the MUCAT Sector at the Advanced 
Photon Source. For the polycrystalline sample, the data were 
integrated over the Debye-Scherrer cone, collected by the area 
detector, to obtain the diffracted intensity as a function of the 
scattering angle 2$\theta$. Additionally, the sample was 
continuously rocked around its vertical axis up to $\pm$2.9~deg. 
An adequate averaging over sample grains was thus achieved. 

FIG.~2 shows patterns measured at base temperature, $T$~=~10~K, 
and at 130~K, which is slightly above the transition temperature 
$T_{\mathrm{t}}$. The lower panel shows the difference between 
the two diffraction patterns. The difference is dominated by the 
decrease of the intensity of Bragg reflections with increasing 
temperature due to thermal motion. Additionally, the thermal 
lattice expansion yields small shifts of the reflections yielding 
differences in the patterns. Beside these effects the diffraction 
patterns remain essentially the same. Therefore, the main 
features of the crystallographic structure are unchanged between 
$T$~=~10~K and 130~K.

\begin{figure}
\includegraphics[width=0.84\linewidth]{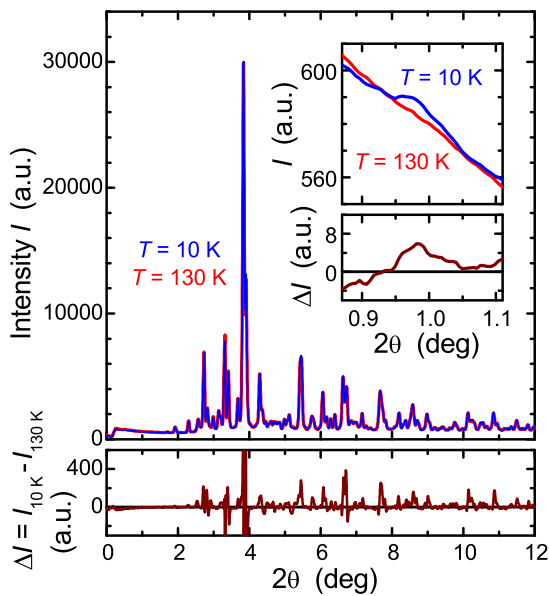}
\caption{\label{fig:fig2} (in color only online) High-energy 
x-ray diffraction patterns of Ce$_{2}$Fe$_{17}$ powder measured 
with $\lambda$~=~0.141~\AA. The upper panel shows diffraction 
patterns at $T$~=~10~K and at 130~K. The difference between the 
intensities at both temperatures is shown in the lower panel. 
Parts of the patterns around the (0~0~3/2) position are enlarged 
in the inset.}
\end{figure}

Below $T_{\mathrm{t}}$, however, a very weak additional 
reflection was observed at low scattering angles, 2$\theta$. The 
inset in FIG.~2 shows this reflection which can be indexed as 
(0~0~3/2). Its intensity increases with decreasing temperature, 
and it is about four orders of magnitude lower than the strongest 
reflections at room temperature, consistent with the 
superstructure observed in the laboratory measurements described 
above. Unfortunately, observations of related superlattice 
reflections at higher scattering angles 2$\theta$ failed due to 
the overlap with much stronger reflections corresponding to the 
room-temperature crystal structure.

\subsection{High-energy x-ray diffraction measurements of a single crystal}

In order to unambiguously determine the nature of the 
superstructure below $T_{\mathrm{t}}$ in Ce$_{2}$Fe$_{17}$, 
measurements of single crystals over a wide range of reciprocal 
space are required. For the single crystal, a special rocking 
technique was applied to record the diffraction intensity within 
planes in reciprocal space. The scattering geometry of FIG.~3 
describes the experiment. In FIG.~3b) the scattering geometry is 
represented by an Ewald sphere fixed in the coordinate system of 
the instrument. The image plate records all points on the Ewald 
sphere, up to the maximum scattering angle 2$\theta$ given by the 
ratio of the dimension of the image plate and the distance 
between sample and detector. The origin of reciprocal lattice is 
located at the center of the recorded pattern. The orientation of 
the reciprocal lattice relative to the Ewald sphere is given by 
the orientation of the sample in the instrument (FIG.~3a)). It 
can be modified by two independent tilting angles perpendicular 
to the incoming beam. If the designated reciprocal plane is 
perpendicular to the incident beam, only one point, the origin, 
is intersected by the Ewald sphere. If the sample is now tilted 
by a small angle, the intersection between the Ewald sphere and 
the designated reciprocal plane is a circle, and can be recorded. 
The diameter of this circle increases with increasing tilting 
angle. In any experimental measurement the resolution is finite, 
and consequently an annulus is recorded as illustrated in 
FIG.~3c). By tilting the sample through both angles, and summing 
the recorded patterns, an extended piece of the designated 
reciprocal plane can be recorded. To obtain the entire reciprocal 
plane without gaps, both tilting angles have to be scanned in a 
two dimensional mesh, and the step size for the sampling mesh has 
to be adjusted to the instrumental resolution and the mosaic 
spread of the sample.

\begin{figure}
\includegraphics[width=0.78\linewidth]{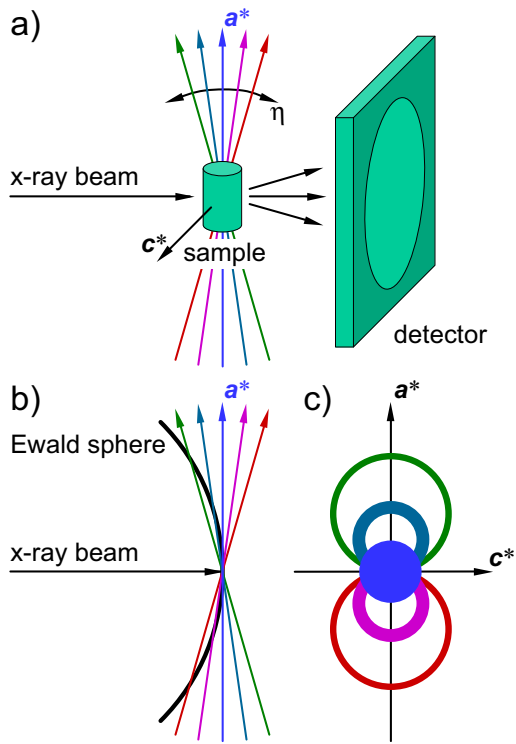}
\caption{\label{fig:fig3} (in color only online) Special rocking 
technique to measure entire reciprocal planes of a 
Ce$_{2}$Fe$_{17}$ single crystal using high-energy x-rays. Here, 
it is illustrated for the recording of the ($H$~0~$L$) plane. 
Dimensions and angles are not to scale and, partly, strongly 
exaggerated for a better illustration. a) Instrumental geometry 
with the image-plate detector mounted centered and perpendicular 
to the incident x-ray beam behind the sample. The direct x-ray 
beam is blocked by a small beam catcher not shown here. Tilting 
by the angle $\eta$ is visualized by the resulting fan of the 
$\mathbf{a}*$ direction. b) Ewald illustration of the scattering 
geometry in reciprocal space. The tilting of the reciprocal 
lattice caused by the sample tilting is visualized by the fan of 
the $\mathbf{a}*$ direction. c) View onto the reciprocal 
($H$~0~$L$) plane in beam direction. The circle and the annuluses 
represent the intersection between the Ewald sphere and the 
($H$~0~$L$) plane for the different tilting angles $\eta$ shown 
in FIG.~1a) and 1b).}
\end{figure}

Unfortunately, points away from the designated reciprocal plane 
are also recorded. Depending on the dimensions of the reciprocal 
lattice and the tilting angles, reciprocal planes other than the 
designated plane will also intersect the Ewald sphere. This 
limits the analyzable area of the recorded pattern and, 
accordingly, the maximum range of meaningful tilting angles. The 
diameter of the analyzable area is proportional to the radius of 
the Ewald sphere given by the length of the wave vector of the 
incident x-rays. The use of high-energy x-rays significantly 
enhances this study because the relatively short wavelength 
corresponds to large a magnitude of the incident wavevector. A 
considerable amount of time was required to read out the image 
plate. Therefore, the image plate was exposed during the full 
time of meshing both tilting angles rather than recording 
separate patterns at stepwise increments in the tilting angles. 
An additional advantage of this method is the opportunity to scan 
one tilting angle continuously, as only the second tilting angle 
needs to be changed stepwise.

The study of the Ce$_{2}$Fe$_{17}$ single crystal was performed by
scanning the horizontal angle $\varphi$ up to $\pm$1.8~deg and
stepping the vertical angle $\eta$ up to $\pm$1.8~deg with a step
size of 0.15~deg. The total exposure time was 150~sec. for each 
pattern, which was determined by the maximum motor speed for the 
tilting angles. Aluminum attenuators were used to adapt the 
scattered intensity to the dynamic range of the image-plate 
system. Furthermore, to eliminate extraneous reflections from the 
Be domes as discussed above, an aluminum enclosure with kapton 
windows was used for these measurements. In FIG.~4 diffraction 
patterns are shown for the three most relevant reciprocal planes 
($H~H~L$), ($H~K~0$), and ($H~0~L$).

\begin{figure}
\includegraphics[width=0.75\linewidth]{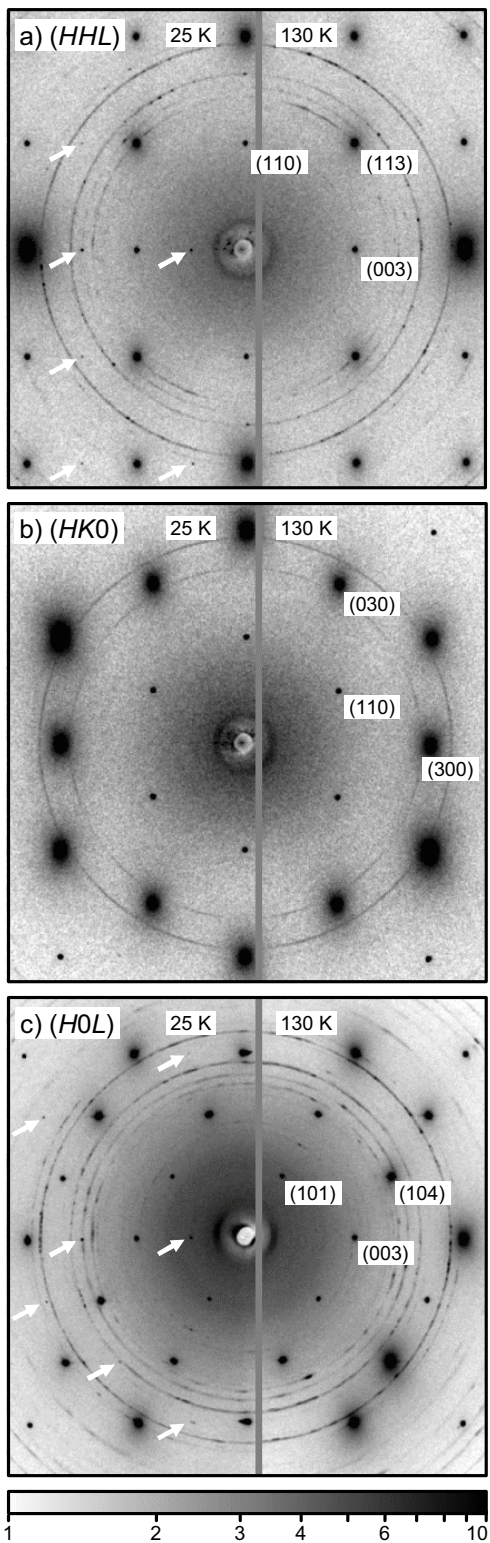}
\caption{\label{fig:fig4} High-energy x-ray diffraction patterns 
of the Ce$_{2}$Fe$_{17}$ single crystal. The intensity is encoded 
by the grey tone in the contour map of the a) ($H~H~L$), b) 
($H~K~0$), and c)($H~0~L$) planes. The left and right side show 
measurements at $T$~=~25~K and at 130~K, respectively, which is 
above the crystallographic phase transition. The arrows point to 
superstructure reflections appearing in the low-temperature 
phase.}
\end{figure}

Above $T_{\mathrm{t}}$, the diffraction patterns were consistent 
with the room-temperature crystal structure. Weak ring-like 
scattering features in the patterns arose from small amounts of 
polycrystalline CeFe$_{2}$ and an undefined polycrystalline 
sample impurity. In the ($H~0~L$) plane, only a two-fold rotation 
axis perpendicular to the diffraction pattern was found as a 
symmetry element. Reflections were observed at positions 
($H~0~L$) with $-H~+~L~=~3n$ where $H$, $L$, and $n$ are 
integers, according to the space group $R~\overline{3}~m$. 
Interestingly, no reflections were found at positions ($H~0~L$) 
with $H~+~L~=~3n$ where $H$, $L$, and $n$ are integers, expected 
for a twinned single crystal. Below $T_{\mathrm{t}}$, 
superstructure reflections were observed in the ($H~0~L$) and 
($H~H~L$) plane at positions displaced by (0~0~3/2) from 
positions related to the room-temperature crystal structure. No 
other extra reflections were observed. In the ($H~K~0$) plane, no 
superstructure reflections were found. All diffraction patterns 
measured at low temperatures are consistent with a unit cell 
doubled along the $c$ direction based on the dimensions of the 
room-temperature unit cell.

\begin{figure}
\includegraphics[width=0.8\linewidth]{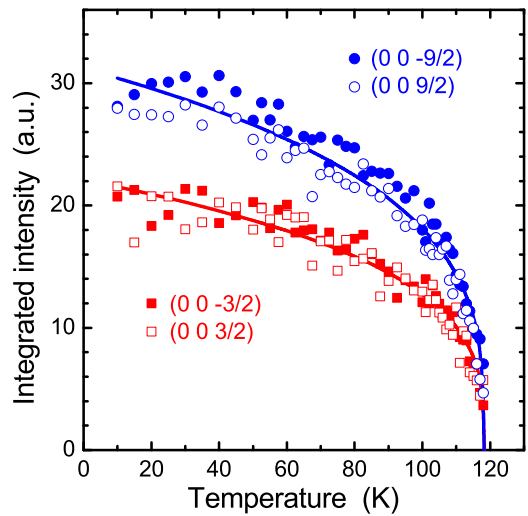}
\caption{\label{fig:fig5} (in color only online) Temperature 
($T$) dependence of the integrated intensity ($I$) of 
superstructure reflections measured with high-energy x-rays on 
the Ce$_{2}$Fe$_{17}$ single crystal. The lines represent 
power-law curves describing the combined data for the 
Friedel-pair reflections (0~0~$\pm$3/2) and (0~0~$\pm$9/2), 
respectively.}
\end{figure}

To measure the temperature dependence of the superlattice 
reflections, diffraction patterns of the ($H~0~L$) plane were 
recorded, but with the sample enclosed by the beryllium domes. 
The reflection intensity $I$ was determined as function of the 
well-defined sample temperature $T$ by fitting two-dimensional 
gaussian-shaped peaks to the diffraction patterns. FIG.~5 shows 
the result for selected superstructure reflections. The good 
agreement between the intensity of corresponding Friedel-pair 
reflections indicates a properly chosen experimental geometry. 
The temperature-dependent data are well described by a power law. 
A transition temperature $T_{\mathrm{t}}$~=~118$\pm$2~K was 
consistently determined for all investigated reflections. This 
transition temperature is consistent with that observed in 
low-field magnetization measurements of the same sample. The 
transition temperature is slightly lower than for the samples in 
Ref.~\cite{Janssen06}. The difference is likely caused by a small 
amount of a dopant. Recently, for example, we have found that a 
doping by only 0.1\% of Al for Fe causes a decrease in 
$T_{\mathrm{t}}$ by 
about 10~K.\cite{Janssen07}\\

\section{Discussion}
\subsection{Superstructure and crystallographic phase transition}

In Ce$_{2}$Fe$_{17}$, a crystallographic superstructure was found
below $T_{\mathrm{t}}$~=~118$\pm$2~K. In addition to the
measurements described here, we have observed similar
crystallographic superstructure reflections in recent neutron
diffraction experiments as well as x-ray resonant magnetic
scattering (XRMS) studies, which will be presented elsewhere.
However, it is worth emphasizing here that the sample penetration
depth for all of these structural probes covers a large range from
$\sim$2~$\mu$m in the XRMS study at the Ce-$L_{2}$ absorption edge,
to 2.2~mm for the high-energy x-ray diffraction experiments, and up
to few cm for the scattering of thermal neutrons, respectively. Thus
these experiments have probed the formation of the superstructure in
different parts of the sample, from near-surface regions through the
entire bulk. In all cases the widths of the superstructure
reflections were similar to the widths of main reflections. This
indicates that the superstructure forms with the same degree of
perfection as the main crystal structure in every part of the
sample. The intensity ratio between the superstructure and main
reflections was consistent in all these studies. From these results
it can be concluded that the superstructure is uniform throughout
the entire Ce$_{2}$Fe$_{17}$ sample.

We now turn to a discussion of the space group of the 
low-temperature superstructure as determined by an analysis of 
possible crystallographic subgroups of the room-temperature space 
group $R~\overline{3}~m$. The observed doubling of the unit cell 
in the hexagonal $c$ direction corresponds to a doubling of the 
length along the body diagonal in the rhombohedral lattice. Only 
two subgroups of the space group $R~\overline{3}~m$ are related 
to a doubling of the unit cell in the $\mathbf{c}$ direction, the 
non-isomorphic subgroup (A) $R~\overline{3}~c$ 
($\mathbf{a}'~=~-\mathbf{a}$, $\mathbf{b}'~=~-\mathbf{b}$, 
$\mathbf{c}'~=~2\mathbf{c}$) and the isomorphic subgroups (B) 
$R~\overline{3}~m$ ($\mathbf{a}'~=~-\mathbf{a}$, 
$\mathbf{b}'~=~-\mathbf{b}$, 
$\mathbf{c}'~=~2\mathbf{c}$).\cite{IntTab96} The prime symbol (') 
denotes the subgroup. Both subgroups (A) and (B) yield a general 
condition for ($H'~K'~L'$) reflections with $-H'~+~K'~+~L'~=~3n$ 
which can be satisfied by the observed diffraction patterns 
considering the transformation $L'/2~=~L$. The subgroup (A) 
$R~\overline{3}~c$ contains an additional general condition for 
($H'~-H'~L'$) reflections with $H'~+~L'~=~3n$ and $L'~=~2n$ 
caused by the glide plane.\cite{IntTab96} However, the observed 
reflections violate this condition, e.g. reflections were present 
at (0~0~3/2) and (0~0~9/2) positions which would be (0~0~3) and 
(0~0~9), respectively, in the notation of the subgroup. As 
result, only subgroup (B) is possible. All observed diffraction 
patterns are in agreement with this subgroup, $R~\overline{3}~m$ 
($\mathbf{a}'~=~-\mathbf{a}$, $\mathbf{b}'~=~-\mathbf{b}$, 
$\mathbf{c}'~=~2\mathbf{c}$) as space group for the 
low-temperature phase.

The doubling of the unit cell in $\mathbf{c}$ relates to a 
substitution of the symmetry element translation with the vector 
$\mathbf{c}$ by a translation with the vector 2$\mathbf{c}$, 
which corresponds to twofold reduction of the symmetry. 
Therefore, the transition from the room-temperature structure to 
the low-temperature superstructure can be described by a 
transition to a direct subgroup of the original space group of 
index two.\cite{IntTab96}

\begin{table}
\begin{center}
\begin{tabular}{p{25pt}ccccp{15pt}cp{2pt}ccc}
\hline\hline
  & room $T$: & $x$ & $y$ & $z$ &  & low $T$: &  & $x$ & $y$ & $z$ \\
\hline
Ce & $6c$ & 0 & 0 & $z$ &  & $6c$ &  & 0 & 0 & $z$/2 \\
 &  &  &  &  &  & $6c$ &  & 0 & 0 & $z$/2+1/2 \\
\hline
Fe & $6c$ & 0 & 0 & $z$ &  & $6c$ &  & 0 & 0 & $z$/2 \\
 &  &  &  &  &  & $6c$ &  & 0 & 0 & $z$/2+1/2 \\
\hline
Fe & $9d$ & 1/2 & 0 & 1/2 &  & $18h$ &  & 0.5 & -0.5 & 0.25  \\
\hline
Fe & $18f$ & $x$ & 0 & 0 &  & $18f$ &  & -$x$ & 0 & 0 \\
 &  &  &  &  &  & $18g$ &  & -$x$ & 0 & 1/2 \\
\hline
Fe & $18h$ & $x$ & -$x$ & $z$ &  & $18h$ &  & -$x$ & $x$ & $z$/2 \\
 &  &  &  &  &  & $18h$ &  & -$x$ & $x$ & $z$/2+1/2 \\
\hline\hline
\end{tabular}
\caption{\label{tab:tab1} Wyckoff sites and atomic coordinates 
for the room-temperature phase in Ce$_{2}$Fe$_{17}$ and their 
transformation into the low-temperature phase. The symbols $x$, 
$y$, $z$ and decimal-point numbers represent free parameters. The 
space group is $R~\overline{3}~m$ for both phases. The lattice 
parameter transform according to 
$\mathbf{a}_{\textrm{low}T}~=~-\mathbf{a}_{\textrm{room}T}$, 
$\mathbf{b}_{\textrm{low}T}~=~-\mathbf{b}_{\textrm{room}T}$, and 
$\mathbf{c}_{\textrm{low}T}~=~2\mathbf{c}_{\textrm{room}T}$.}
\end{center}
\end{table}

TABLE~1 lists the transformation of the atomic positions and their
Wyckoff sites according to the subgroup description defined above.
For atoms on the original Wyckoff sites 6$c$, 18$f$, and 18$h$, 
the site symmetry is not changed by the phase transition. 
Consequently, these sites split up into two new sites with a 
doubling of the number of free parameters for the atomic 
positions according to the index two of the subgroup. The 
situation is different for the original Wyckoff site 9$d$ which 
transforms to the new Wyckoff site 18$h$. The new site symmetry 
$.m$ is, by index two, lower than the original site symmetry 
$.2/m$. The missing symmetry element yields two new free 
parameters for the new Wyckoff position, ($x,~-x,~z$), in 
contrast to the fixed original Wyckoff position, 
($1/2,~0,~1/2$),\cite{IntTab96} see TABLE~1.

The observed intensity of superstructure reflections can be used 
for an order-of-magnitude estimate of the corresponding 
displacement of atoms from their room-temperature position with 
higher symmetry. The observed intensity ratio between the 
superstructure reflection (0~0~15/2) at $T$~=~110~K and the main 
reflection (0~0~6) at $T$~=~130~K, respectively, is roughly 
4~:~9500 as shown in FIG.~1. This intensity ratio depends only on 
the $z$ position of the atoms because (0~0~$L$) reflections 
average structure features perpendicular to the $z$ direction. 
Two different cases of atomic displacements are considered. In 
the first, the superstructure is caused by a displacement of only 
one atomic site, the Fe site $6c$, for example. A relative 
displacement of $\pm$0.005 r.l.u. along the $z$ direction is 
required to obtain the observed intensity ratio. In a second 
scenario, all atoms are displaced in $z$ direction by an equal 
distance. The sense of direction for the displacement is chosen 
in such a manner that the intensity ratio is maximized. The 
result is a relative displacement of $\pm$0.0002 r.l.u. These two 
cases represent roughly the upper and lower limit for atomic 
displacements related to the observed superstructure. These very 
slight changes in the crystal structure are consistent with the 
observed marginal changes of the main reflections.

According to Landau's theory, a phase transition described by a 
subgroup relation is consistent with a second order phase 
transition where a symmetry element is broken and an order 
parameter starts to develop.\cite{Landau37} The atomic 
displacement must follow the temperature dependence of the 
intensity, because the intensity of superstructure reflections 
depends quadratically on the value of the atomic displacement. 
Hence the atomic displacement can be considered as the order 
parameter of this phase transition. It starts at the transition 
temperature without a jump. Therefore, no abrupt re-arrangement 
of atoms is necessary to originate the observed behavior around 
the phase transition. It can be described as a diffusion-less 
movement of atoms consistent with a second order phase transition.

\subsection{Influence of the crystallographic phase transition on magnetic and electronic properties}

As discussed in the introduction, the magnetic properties of 
Ce$_{2}$Fe$_{17}$ seem particularly sensitive to small changes in 
chemistry and structure.  The magnetic and electronic properties 
can be influenced in several manners by the observed 
crystallographic phase transition. We now consider the effect of 
the change in local symmetries and of the additional freedom in 
atomic positions as well as the consequences of the doubling of 
the unit cell.

A lowering of the site symmetry is associated with a lowering of 
the symmetry of the crystal-electric field for the atoms on this 
site. However, only minor changes in the magneto-crystalline 
anisotropy are expected because crystal-electric field effects 
are less important in Ce$_{2}$Fe$_{17}$. More relevant for the 
magnetism, is the splitting of original single Wyckoff sites into 
pairs of Wyckoff sites. The related doubling of the number of 
free parameters for the atomic positions yields changes in the 
distance and angle between magnetic neighbor atoms which can, in 
principle, affect the strength of the magnetic 
exchange.\cite{Coehoorn89,Givord74} This can be important for 
Ce$_{2}$Fe$_{17}$ in light of the delicate balance between 
antiferromagnetic and ferromagnetic interactions. For example, 
for the Fe atoms originally located on the Wyckoff site 9$d$, the 
effect is not only quantitative, it is qualitative due to the 
change in the local symmetry at the phase transition. For the 
room-temperature structure, these atoms are located on a two-fold 
rotation axis. Thus, neighbor atoms are pairwise 
symmetry-coupled, and the distances between them are equal. This 
condition is lost in the low-temperature structure, and the 
distances between these neighboring atoms are different. 
Consequently, the strength of the magnetic exchange between these 
neighboring atoms is no longer equal, possibly leading, in some 
cases, to a change in sign of the magnetic exchange.

In the low-temperature phase, the doubling of the unit cell along 
the hexagonal $\mathbf{c}$ direction in real space halves the 
Brillouin zone in $\mathbf{c}^{*}$ direction, and the shape of 
the Brillouin zone changes accordingly, because the dimension in 
$\mathbf{c}^{*}$ direction is now much smaller than in other 
directions. This is evidenced in FIG.~4 by the smaller distances 
along the $\mathbf{c}^{*}$ direction between reflections, normal 
and superstructure reflections together, in comparison to other 
directions. Modifications of the electronic band structure, 
associated with the opening of a new superzone gap, can be 
expected and can result in changes in the transport properties. 
Indeed, in measurements of the electric conductivity on 
Ce$_{2}$Fe$_{17}$ single crystals, a strong increase in 
resistivity was observed below the transition temperature 
$T_{\mathrm{t}}$.\cite{Janssen06} This increase is strongly 
anisotropic; it is much larger for the current flowing parallel 
to the $\mathbf{c}$ direction than for the current flowing 
perpendicular to the $\mathbf{c}$ direction. A detailed 
description of this effect consistent with the formation of 
anisotropic superzone gaps is given in Ref.~\cite{Janssen06}. It 
should be mentioned, that the crystallographic phase transition 
occurs in the antiferromagnetically ordered state with an 
incommensurate propagation vector in $\mathbf{c}^{*}$ direction. 
Both superstructures, the crystallographic superstructure and the 
antiferromagnetic superstructure, can yield superzone gaps and, 
therefore, can cause the change in resistivity at the phase 
transition temperature $T_{\mathrm{t}}$. In break-junction 
measurements, the differential conductance exhibits gap features 
associated with the low-temperature phase below 
$T_{\mathrm{t}}$.\cite{Ekino99} Remarkably, features were 
determined at two energy scales 12-14~meV and 26-32~meV on a 
polycrystalline sample at $T$~=~4.2~K. This could be related to 
the existence of two different gap features probably caused by 
the changes in the crystal structure and in the magnetic order.

In addition to generating superlattice reflections, the change in 
the chemical unit cell below $T_{\mathrm{t}}$ also modifies the 
description of the magnetic order in reciprocal space and, 
accordingly, the pattern for magnetic scattering. In neutron 
powder experiments, satellite reflections related to the 
antiferromagnetic structure with an incommensurate propagation 
vector in the hexagonal $\mathbf{c}^{*}$ direction were observed 
below $T_{\mathrm{N}}$.\cite{Plumier74,Fukuda99,Fukuda99a} Below 
$T_{\mathrm{t}}$ additional satellite reflections were observed, 
which could be indexed with the same propagation vector, but 
starting from the reference position (0~0~3/2). The crystal 
superstructure determined in the present publication yield new 
lattice points at just such positions. Therefore, the results of 
the earlier neutron scattering 
experiments\cite{Plumier74,Fukuda99,Fukuda99a} can be 
reinterpreted as the observation of magnetic satellite 
reflections according to the same incommensurate propagation 
vector based upon the lattice of the new superstructure. 
Accordingly, the number of pairs of observed antiferromagnetic 
satellite reflections is doubled below $T_{\mathrm{t}}$ in 
comparison to the number above $T_{\mathrm{t}}$.

Based on these results, additional information can now be 
extracted from the neutron powder diffraction patterns of 
Ce$_{2}$Fe$_{17}$ measured under isotropic pressures up to 
5~kbar.\cite{Prokhnenko02a,Prokhnenko02,Prokhnenko04} At ambient 
pressure, a ferromagnetic state was observed in this sample. 
Possible reasons for this different magnetic state are related to 
sample preparation and discussed in Ref.~\cite{Janssen06}. An 
antiferromagnetic state can be induced by pressures higher than 
3~kbar at $T$~=~40~K.\cite{Prokhnenko02a} Two groups of 
antiferromagnetic satellite reflections appear. The first group 
consists of satellite reflections related to an incommensurate 
propagation vector similar to those observed for 
antiferromagnetic order at higher temperatures at ambient 
pressure. The second group of reflections were indexed with a 
new, different incommensurate propagation 
vector.\cite{Prokhnenko02a} However, these new satellite 
reflection at 2$\theta$~$\sim$~11~deg. shown in FIG.~11 of 
Ref.~\cite{Prokhnenko02a} can be indexed by (0~0~3/2)$^{-}$. All 
new satellite reflections can be indexed in the same way as the 
low-temperature antiferromagnetic phase discussed in the present 
work. This reinterpretation is supported by the pressure 
dependence of the positions of these reflections, the 
(0~0~3/2)$^{-}$ reflection moves in the opposite direction to the 
(0~0~0)$^{+}$ reflection. This suggests that the 
ferromagnetic-antiferromagnetic phase transition at a pressure of 
3~kbar is connected with the appearance of a similar 
superstructure as observed in Ce$_{2}$Fe$_{17}$ samples, which 
show an antiferromagnetic state at low temperatures at ambient 
pressure.

\section{Conclusions}

In conclusion, a superstructure was observed in 
Ce$_{2}$Fe$_{17}$, which appears below 
$T_{\mathrm{t}}$~=~118$\pm$2~K in the antiferromagnetically 
ordered state. It can be described by a doubling of the unit cell 
in hexagonal $\mathbf{c}$ direction and the same space group 
$R~\overline{3}~m$ as the room-temperature structure. The second 
order phase transition is realized by a diffusion-less movement 
of atoms. The additional freedom in atomic positions and the 
change in local symmetry for the Fe atoms at the original Wyckoff 
sites 9$d$ can influence the magnetic behavior due to the  
delicate balance between antiferromagnetic and ferromagnetic 
interactions which are strongly dependent on the distance between 
neighbor Fe atoms. The halving of the Brillouin zone in 
$\mathbf{c}^{*}$ direction can cause a superzone gap in the 
electronic structure, which is consistent with observations of an 
anisotropic increase in the electric resistivity below 
$T_{\mathrm{t}}$. Additional antiferromagnetic satellite 
reflections observed in former neutron diffraction experiments 
can be conclusively related to the appearance of the 
crystallographic superstructure.

\section{ACKNOWLEDGEMENTS}

The authors thank J. Frederick and S. Jia for the assistance in 
sample preparation. Ames Laboratory is supported by the 
Department of Energy, Office of Science under Contract No. 
W-7405-ENG-82. Use of the Advanced Photon Source (APS) is 
supported by the U.S. Department of Energy, Basic Energy 
Sciences, Office of Science, under Contract No. W-31-109-ENG-38. 
The Midwest Universities Collaborative Access Team sector at the 
APS is supported by the Department of Energy, Office of Science 
through the Ames Laboratory Contract No. W-7405-ENG-82.

\end{document}